\documentclass[12pt]{article}

\begin{document}

\title{Gentile statistics with a large maximum occupation number}
\author{Wu-Sheng Dai$^{1),3)}$ \thanks{%
Email: daiwusheng@tju.edu.cn} and Mi Xie$^{2)}$ \thanks{%
Email: xiemi@mail.tjnu.edu.cn} \\
{\footnotesize 1) School of Science, Tianjin University, Tianjin 300072, P.
R. China}\\
{\footnotesize 2) Department of Physics, Tianjin Normal University, Tianjin
300074, P. R. China}\\
{\footnotesize 3) LiuHui Center for Applied Mathematics, Nankai University
\& Tianjin University,}\\
{\footnotesize Tianjin 300072, P. R. China} }
\date{}
\maketitle

\begin{abstract}
In Gentile statistics the maximum occupation number can take on unrestricted
integers: $1<n<\infty $. It is usually believed that Gentile statistics will
reduce to Bose-Einstein statistics when $n$ equals the total number of
particles in the system $N$. In this paper, we will show that this statement
is valid only when the fugacity $z<1$; nevertheless, if $z>1$ the
Bose-Einstein case is not recovered from Gentile statistics as $n$ goes to $%
N $. Attention is also concentrated on the contribution of the ground state
which was ignored in related literature. The thermodynamic behavior of a $%
\nu $-dimensional Gentile ideal gas of particle of dispersion $E=\frac{p^{s}%
}{2m}$, where $\nu $ and $s$ are arbitrary, is analyzed in detail. Moreover,
we provide an alternative derivation of the partition function for Gentile
statistics.
\end{abstract}


\section*{I. Introduction}

Quantum statistics is classified as either Bose-Einstein or Fermi-Dirac
statistics. As a generalization, Gentile \cite{Gentile} proposed a kind of
intermediate statistics, in which the maximum number of particles in any
quantum state is neither $1$ nor $\infty $, but equal to a finite number $n$%
. A particularly interesting case of Gentile statistics is $n\rightarrow N$,
where $N$ is the total number of particles in the system. Some authors,
e.g., Ref. \cite{ter Haar}, argue that in this case there is no difference
between Bose-Einstein and Gentile statistics. However, this statement is
valid only when the fugacity $z<1$. In the present paper, we will discuss
the case of $n\rightarrow N$ with $z>1$ and show that in this case the
Bose-Einstein case can not be recovered from Gentile statistics.

The thermodynamic behavior of an ideal gas obeying Gentile statistics has
been discussed in some literature \cite{Gentile,Katsura}. Nevertheless, in
these works the contribution from the ground state was ignored. It is known
that such a contribution can be safely ignored in fermion systems; however,
in boson systems it plays an important role especially in the case of low
temperatures and high densities. In Gentile statistics, the maximum
occupation number $n$ can take on unrestricted integers: $1<n<\infty $. When
$n$ is large, especially $n\rightarrow N$ , we will show that the
contribution from the ground state becomes very important.

In the present paper, we will calculate the thermodynamic quantities in $\nu$%
-dimensional space with the dispersion relation $E=\frac{p^{s}}{2m}$, where $%
\nu $ and $s$ are arbitrary. A special case of $\nu =3$ and $s=2$, without
the ground state contribution, has been discussed in Ref. \cite%
{Gentile,Katsura}. Our result shows that, the thermodynamic behavior of
Gentile statistics, especially the contribution of the ground state for the
case $\nu >s$, depends sensitively on the maximum occupation number $n$.

In Sec. II, we first provide an alternative way to calculate the partition
function for Gentile statistics. Special attention is concentrated on the
case of $n\rightarrow N$ with $z>1$. In this case, Gentile statistics will
not reduce to Bose-Einstein statistics. In Sec. III, we calculate the
thermodynamic quantities in which the contribution from the ground state is
reckoned in. In Sec. IV, we discuss the low temperature and high density
result for the internal energy, the specific heat, the magnetic
susceptibility and other thermodynamic quantities. In this case, the ground
state contribution plays an important role. A summary and conclusions are
given in Sec. V.

\section*{II. The partition function and the case of $n\rightarrow N$ with $%
z>1$ in Gentile statistics}

For completeness, we first calculate the partition function for Gentile
statistics in a way which is somewhat different from that in literature. The
approach we shall employ is based on multinomial theorem.

Starting with counting the number of states, we express the grand partition
function as follows:

\begin{equation}
\Xi =\displaystyle\prod\limits_{\ell }\Xi _{\ell }=\displaystyle%
\prod\limits_{\ell }\sum\limits_{a_{\ell }}\Omega _{\ell }e^{-(\alpha +\beta
\varepsilon _{\ell })a_{\ell }}.
\end{equation}%
Here
\begin{equation}
\Xi _{\ell }=\sum\limits_{a_{\ell }}\Omega _{\ell }e^{-(\alpha +\beta
\varepsilon _{\ell })a_{\ell }},  \label{e3.0a}
\end{equation}%
and $\Omega _{\ell }$ is the number of quantum states of $a_{\ell }$
particles distributed in one level $\varepsilon _{\ell }$ (the ${\ell }$-th
level) with a degeneracy $\omega _{\ell }$. In this occasion, one state can
be occupied by at most $n$ particles. It means
\begin{equation}
\Omega _{\ell }=\sum\limits_{\alpha _{0}}\cdots \sum\limits_{\alpha _{n}}%
\frac{\omega _{\ell }!}{\prod\limits_{i=0}^{n}\alpha _{i}!},  \label{e3.1}
\end{equation}%
where $\alpha _{i}~(i=0,1,\cdots ,n)$ denotes the number of states which
contains $i$ particles, and the summation runs over all possible values. The
set of $\{\alpha _{i}\}$ must satisfy the conditions
\begin{equation}
\sum\limits_{i}\alpha _{i}=\omega _{\ell },~~~~~\sum\limits_{i}i\alpha
_{i}=a_{\ell }.  \label{e3.1a}
\end{equation}

To obtain the grand partition function, we need the multinomial theorem \cite%
{Abramowitz}: The coefficient $\Omega $ of $x^{a}~(0\leq a\leq n\omega )$ in
the expansion of $(1+x+x^{2}+\cdots +x^{n})^{\omega }$ is
\[
\Omega =\sum\limits_{\alpha _{0}}\cdots \sum\limits_{\alpha _{n}}\frac{%
\omega !}{\prod\limits_{i=0}^{n}\alpha _{i}!},
\]%
where $\alpha _{i}~(i=0,1,\cdots ,n)$ satisfy $\sum\limits_{i=0}^{n}\alpha
_{i}=\omega $, $\sum\limits_{i=0}^{n}i\alpha _{i}=a$, and the summation runs
over all possible values.

Based on such a relation, we can obtain the analytic form of the $\Xi _{\ell
}$. The $\Omega _{\ell }$ in Eq.(\ref{e3.1}) (constrained by Eq.(\ref{e3.1a}
)) is the coefficient of the term $e^{-(\alpha +\beta \varepsilon _{\ell
})a_{\ell }}$ in the expansion of
\begin{equation}
\left[ 1+e^{-(\alpha +\beta \varepsilon _{\ell })}+e^{-2(\alpha +\beta
\varepsilon _{\ell })}+\cdots +e^{-n(\alpha +\beta \varepsilon _{\ell })}%
\right] ^{\omega _{\ell }}.  \label{e3.2}
\end{equation}%
As a result, Eq.(\ref{e3.0a}) can be calculated exactly:
\begin{equation}
\Xi _{\ell }=\left[ \frac{e^{-(n+1)(\alpha +\beta \varepsilon _{\ell })}-1}{%
e^{-(\alpha +\beta \varepsilon _{\ell })}-1}\right] ^{\omega _{\ell }}.
\label{e3.3}
\end{equation}%
The result agrees with that found by Gentile \cite{Gentile}.

Obviously, when $n=\infty $ the necessary condition on obtaining Eq.(\ref%
{e3.3}) from Eq.(\ref{e3.2}) is the fugacity $z<1$ ($z=e^{-\alpha }$);
otherwise, $\Xi _{\ell }$ will diverge. This means that Gentile statistics
will return to Bose case only when $z<1$. Many authors, e.g., Ref. \cite{ter
Haar}, however, argue that it does not make any difference whether one take $%
n=N$ or $n=\infty $; in other words, Bose-Einstein statistics is recovered
when $n=N$. Nevertheless, Eq.(\ref{e3.3}) will hold so long as $n\neq \infty
$ even for $n\gg 1$ and $z>1$. The behavior of the statistics corresponding
to $z>1$ and $n\gg 1$ will be very different from Bose-Einstein statistics.
In Sec. IV, we will discuss such statistics in detail.

\section*{III. Thermodynamics with the contribution of the ground state}

In this section, we will discuss the thermodynamics corresponding to Gentile
statistics in detail. Specifically, we will study the contribution of the
ground state to the thermodynamic quantities, which is ignored in Ref. \cite%
{Gentile,Katsura}.

The grand partition function for Gentile statistics is
\begin{equation}
\Xi =\displaystyle\prod\limits_{\ell }\left[ \frac{z^{n+1}e^{-(n+1)\beta
\varepsilon _{\ell }}-1}{ze^{-\beta \varepsilon _{\ell }}-1}\right] ^{\omega
_{\ell }},
\end{equation}%
Replacing the summation over $l$ by the corresponding integral and carrying
out the integral by parts, we have
\begin{eqnarray}
\ln \Xi &=&\displaystyle\sum_{l}\ln \Xi _{l}  \nonumber \\
&=&\displaystyle\frac{1}{\lambda ^{\nu }}\frac{2gV}{\nu \Gamma (\frac{\nu }{2%
})}\int_{0}^{\infty }\left[ \frac{1}{z^{-1}e^{\xi }-1}-\frac{n+1}{%
z^{-(n+1)}e^{(n+1)\xi }-1}\right] \xi ^{\frac{\nu }{s}}d\xi +\ln \frac{%
1-z^{n+1}}{1-z},~  \label{e3.8}
\end{eqnarray}%
while we have used the asymptotic formula in $\nu $ dimensions with the
dispersion relation $E=\frac{p^{s}}{2m}$:

\[
\sum_{l}\longrightarrow g\frac{V}{h^{\nu }}\int \frac{2\pi ^{\nu /2}}{\Gamma
(\frac{\nu }{2})}\frac{1}{s}(2m)^{\frac{\nu }{s}}\epsilon ^{\frac{\nu }{s}%
-1}d\epsilon ,
\]%
and introduced $\epsilon =kT\xi $. It should be emphasized that, the
parameter $m$ in the dispersion relation can be explained as the mass of
particles only when $s=2$. Here $\lambda $ is the thermal wavelength

\begin{equation}
\lambda =\left( \frac{h^{s}}{2\pi ^{s/2}mkT}\right) ^{1/s},
\end{equation}%
and $g$ is a weight factor that arises from the internal structure of the
particles (the number of internal degrees of freedom). We have split off the
term in Eq.(\ref{e3.8}) corresponding to $\epsilon =0$, which we explicitly
retain because, when $\nu >s$, the density of states is zero. The physical
interpretation of such a term is similar to the Bose case: Sometimes the
single term $\epsilon =0$ may be as important as the entire sum. The term $%
\epsilon =0$ represents the contribution of the ground state of the system.
The contribution of a state is proportional to the number of particles which
are accommodated in the state. As we know, particles tend to occupy the
lower state as possible as they can. At the low temperature limit the
particles all want to occupy the ground state; thus the particle number of
the ground state nearly equals to the maximum occupation number of the
state. Therefore, when $T\rightarrow 0$ the contribution of the ground state
will be proportional to the maximum occupation number $n$. For fermions, the
contribution from the ground state can be ignored since $n=1$; however, for
bosons, $n=\infty $, the contribution from the ground state becomes very
important in the low-temperature limit. Indeed, when $\nu >s$, the term $%
\epsilon =0$ diverges in the Bose case and causes the phase
transition-----Bose-Einstein condensation. In Gentile statistics the maximum
occupation number $n$ is finite, so the term $\epsilon =0$ does not diverge
and there are no phase transitions like Bose-Einstein condensation. However,
when $n$ is large the influence of such a term may be very important. In the
following we will show this influence on the thermodynamic quantities.
Moreover, to obtain the result corresponding to $\nu \leq s$, we only need
to drop the term $\epsilon =0$.

In the theory of Bose-Einstein and Fermi-Dirac systems we come across
integrals of the type

\[
g_{\sigma }(z)=\frac{1}{\Gamma (\sigma )}\int_{0}^{\infty }\frac{x^{\sigma
-1}}{z^{-1}e^{x}-1}dx\mbox{\ \ \ \ and\ \ \ \ \ }f_{\sigma }(z)=\frac{1}{%
\Gamma (\sigma )}\int_{0}^{\infty }\frac{x^{\sigma -1}}{z^{-1}e^{x}+1}dx.
\]%
To deal with the systems with the grand potential Eq.(\ref{e3.8}), we can
introduce the function

\begin{equation}
h_{\sigma }(z)=\frac{1}{\Gamma (\sigma )}\int_{0}^{\infty }\left[ \frac{1}{
z^{-1}e^{x}-1}-\frac{n+1}{z^{-(n+1)}e^{(n+1)x}-1}\right] x^{\sigma -1}dx.
\label{e3.9}
\end{equation}
The function $h_{\sigma }(z)$ will return to Bose-Einstein or Fermi-Dirac
integrals when $n$ equals to $1$ or $\infty $, respectively. A simple
process of differentiation brings out the relationship between $h_{\sigma
}(z)$ and $h_{\sigma -1}(z)$:

\begin{equation}
h_{\sigma -1}(z)=z\frac{\partial }{\partial z}h_{\sigma }(z).
\end{equation}%
In terms of the function $h_{\sigma }(z)$ we can express the grand potential
Eq.(\ref{e3.8}) in the following form:

\begin{equation}
\ln \Xi =\sum_{l}\ln \Xi _{l}=\frac{1}{\lambda ^{\nu }}\frac{2gV}{\nu \Gamma
(\frac{\nu }{2})}\Gamma (\frac{\nu }{s}+1)h_{\frac{\nu }{s}+1}(z)+\ln \frac{%
z^{n+1}-1}{z-1}.  \label{e3.10}
\end{equation}%
The last term describes the contribution from the ground state. For bosons,
this term returns to $-\ln (1-z)$ with $n\rightarrow \infty $ and $z<1$,
which is in connection with the Bose-Einstein condensation; for fermions,
this term may be ignored since the ground state of a fermion system can
contain only one particle. In Gentile statistics $1<n<$ $\infty $, the
contribution of the ground state depends on the value of $n$. When $n$ is
close to $1$, like the Fermi case, the contribution of the ground state can
be ignored; however, for $\nu >s$, when $n$ is comparable in magnitude to
the total number of particles, the ground state contribution becomes
important. Notice that, although there are no divergences, consequently
there are no phase transitions, the influence which is related to $n$ also
can be very large. Ideally, phase transitions display only when infinities
appear. In practice, however, infinity only means very large, so when $%
n\rightarrow N$, though there are no divergences, the behavior of the system
will be very similar to phase transitions.

Since the grand potential Eq.(\ref{e3.10}) involves the contribution
corresponding to $\epsilon =0$, from here onward, the thermodynamic
quantities which reckon in the influence of the ground state follow
straightforwardly.

Eliminating $z$ between the two equations

\begin{equation}
\left\{
\begin{array}{l}
\displaystyle\frac{P}{kT}=\frac{1}{V}\ln \Xi =\frac{2g}{\lambda ^{\nu }}%
\frac{\Gamma (\frac{\nu }{s}+1)}{\nu \Gamma (\frac{\nu }{2})}h_{\frac{\nu }{s%
}+1}(z)+\frac{1}{V}\ln \frac{z^{n+1}-1}{z-1}, \\
\displaystyle\frac{1}{v}=\frac{N}{V}=\frac{1}{V}z\frac{\partial }{\partial z}%
\ln \Xi =\frac{2g}{\lambda ^{\nu }}\frac{\Gamma (\frac{\nu }{s}+1)}{\nu
\Gamma (\frac{\nu }{2})}h_{\frac{\nu }{s}}(z)+\frac{N_{0}}{V},%
\end{array}%
\right.  \label{e18.2}
\end{equation}%
we obtain the equation of state of ideal Gentile gases. Here, the last terms
of these two equations describe the contribution from the ground state, and

\begin{equation}
N_{0}=n-(\frac{1}{z-1}-\frac{n+1}{z^{n+1}-1})
\end{equation}%
is the average occupation number of the ground state. Easily to see that
when $n=\infty $ and $z<1$ the occupation number $N_{0}=\frac{z}{1-z}$. This
is just the case of bosons. In the case of fermions, $N_{0}=\frac{z}{1+z}$
and its contribution is negligible compared to the first term in any cases.
In Gentile statistics $1<n<\infty $, the influence of such a term depends on
the value of $n$. When $n$ is large this influence becomes important.

The internal energy of the ideal gas is given by

\begin{equation}
\frac{U}{N}=-\frac{1}{N}\frac{\partial }{\partial \beta }\ln \Xi =(1-\frac{%
N_{0}}{N})\frac{\nu }{s}kT\frac{h_{\frac{\nu }{s}+1}(z)}{h_{\frac{\nu }{s}%
}(z)}.
\end{equation}%
For the Helmholtz free energy of the gas, we get

\begin{equation}
F=N\mu -PV=NkT\left[ \ln z-(1-\frac{N_{0}}{N})\frac{h_{\frac{\nu }{s}+1}(z)}{%
h_{\frac{\nu }{s}}(z)}-\frac{1}{N}\ln \frac{z^{n+1}-1}{z-1}\right] ,
\end{equation}
and for the entropy

\begin{equation}
S=\frac{U-F}{T}=Nk\left[ (1-\frac{N_{0}}{N})(\frac{\nu }{s}+1)\frac{h_{\frac{%
\nu }{s}+1}(z)}{h_{\frac{\nu }{s}}(z)}-\ln z+\frac{1}{N}\ln \frac{z^{n+1}-1}{%
z-1}\right] .
\end{equation}
The specific heat capacity is given by

\begin{eqnarray}
C_{V} &=&\frac{\partial U}{\partial T} =\frac{\nu }{s}Nk(1-\frac{N_{0}}{N}) %
\left[ \ (\frac{\nu }{s}+1)\frac{h_{ \frac{\nu }{s}+1}(z)}{h_{\frac{\nu }{s}%
}(z)} \right.  \nonumber \\
& & \left. -\frac{\nu }{s}\left( 1+\frac{1}{N-N_{0}}\left( \frac{z}{(z-1)^{2}%
}-\frac{(n+1)^{2}z^{n+1}}{(z^{n+1}-1)^{2}}\right) \frac{h_{\frac{\nu }{s}}(z)%
}{h_{\frac{\nu }{s}-1}(z)}\right) ^{-1}\frac{h_{\frac{\nu }{s}}(z)}{h_{\frac{%
\nu }{s}-1}(z)}\right] .
\end{eqnarray}%
In the derivation of $C_{V}$ we have used the relation

\begin{equation}
\frac{\partial z}{\partial T}=-\frac{\nu }{s}\frac{z}{T}\frac{h_{\frac{\nu }{%
s}}(z)}{h_{\frac{\nu }{s}-1}(z)}\left[ 1+\frac{1}{N-N_{0}}\left( \frac{z}{%
(z-1)^{2}}-\frac{(n+1)^{2}z^{n+1}}{(z^{n+1}-1)^{2}}\right) \frac{h_{\frac{%
\nu }{s}}(z)}{h_{\frac{\nu }{s}-1}(z)}\right] ^{-1},
\end{equation}%
which can be obtained from the expression of $N$ in Eq.(\ref{e18.2}). The
existence of ground state term makes the formulas somewhat complex.

Comparing with the thermodynamic quantities given by Ref. \cite%
{Gentile,Katsura}, we can see that there is an extra factor $1-\frac{N_{0}}{N%
}$ in our result. The addition of such a factor to the thermodynamic
quantities brings in important information: the influence of the ground
state. If we ignore the contribution from the ground state or, equivalently,
take $N_{0}=0$, the above thermodynamic quantities will return to the
results given by Ref. \cite{Gentile,Katsura}. However, since in Gentile
statistics $N_{0}$ can take any value, the influence of the ground state
must be reckoned in. For instance, when $z\gg 1$ and $N_{0}\approx N$ so
that the factor $1-\frac{N_{0}}{N}\rightarrow 0$, the thermodynamic
quantities, such as the internal energy, will approach $0$. The ratio $\frac{%
N_{0}}{N}$ reflects the difference of the various intermediate statistics
corresponding to different $n$. Notice that, the factor $1-\frac{N_{0}}{N}$
comes from the term corresponding to $\epsilon =0$. Such a term is needed
only when $\nu >s$.

For studying the properties of thermodynamic quantities in detail, we first
need to discuss the behavior of $z$ as determined by the second equation of
Eq.(\ref{e18.2}), namely:

\begin{equation}
\frac{\lambda ^{\nu }}{v}=\frac{1}{1-\frac{N_{0}}{N}}2g\frac{\Gamma (\frac{%
\nu }{s}+1)}{\nu \Gamma (\frac{\nu }{2})}h_{\frac{\nu }{s}}(z).
\end{equation}%
Instead of the Bose-Einstein and the Fermi-Dirac integrals $g_{\frac{\nu }{s}%
}(z)$ and $f_{\frac{\nu }{s}}(z)$, there is a function $h_{\frac{\nu }{s}%
}(z) $. To determine the range of the fugacity $z$, we rewrite the factor in
function $h_{\frac{\nu }{s}}(z)$ in the following form:

\begin{equation}
\frac{1}{\xi -1}-\frac{n+1}{\xi ^{n+1}-1}=\frac{\xi ^{n-1}+2\xi
^{n-2}+\cdots +n}{\xi ^{n}+\xi ^{n-1}+\cdots +1},  \label{e18.8}
\end{equation}%
where $\xi =z^{-1}e^{x}$. It is obvious that for any real values of $z$, $h_{%
\frac{\nu }{s}}(z)$ is a bounded, positive function of $z$.

From Eq.(\ref{e18.8}), one can learn that if we expect $h_{\frac{\nu }{s}%
}(z) $ to return to the Bose-Einstein integral the necessary condition $z<1$
is needed so that $h_{\frac{\nu }{s}}(z)$ is bounded. In other words, if we
expect the above formulas to return to the Bose case when $n=\infty $, we
need an additional restriction $z<1$. For comparison we recall that $0\leq
z<\infty $ in Fermi-Dirac statistics and Gentile statistics.

\section*{IV. Low temperatures and high densities}

In this section, we will focus on the behavior of the statistics
corresponding to $z\gg 1$. For $\lambda ^{\nu }/v\gg 1$ the average thermal
wavelength is much greater than the average interparticle separation. It is
easy to see that this is equivalent to the requirement $z\gg 1$. It can be
expected that in this case the contribution from the ground state will
become important. It is noteworthy that, in this case, the result will not
return to the Bose case even when $n\rightarrow N$. To obtain Bose-Einstein
statistics from Gentile statistics, one needs to not only perform the limit $%
n\longrightarrow N$, but also restrict the fugacity $z\leq 1$. Thus the {%
asymptotic }results of the thermodynamic functions for low temperatures and
high densities given in this section will not return to the Bose case since
the asymptotic expansion is based on $z\gg 1$. (We need not to consider the
case of $n\rightarrow N$ with $z<1$ because this is just the Bose case \cite%
{ter Haar}.)

To study the behavior of $h_{\sigma }(z)$, which contains two integrals, for
large $z$, we introduce the variable $t=\ln z$ so that the first integral in
Eq.(\ref{e3.9}) can be expanded in powers of $t$:

\begin{eqnarray}
I &=&\int_{0}^{\infty }\frac{\xi ^{\frac{\nu }{s}}}{z^{-1}e^{\xi }-1}d\xi
\nonumber \\
&=&\int_{0}^{\infty }\frac{\xi ^{\frac{\nu }{s}}}{e^{\xi -t}-1}d\xi
\nonumber \\
&\simeq &-\frac{t^{\frac{\nu }{s}+1}}{\frac{\nu }{s}+1}+2\sum_{j=1,3,5\cdots
}\left( \matrix{ \frac{\nu }{s} \cr j}\right) t^{\frac{\nu }{s}-j}\Gamma
(1+j)\zeta (1+j),
\end{eqnarray}
while the lower limit $t$ in one of the integral is approximately replaced
by $\infty $. By the same procedure, we can derive the second integral:

\begin{eqnarray}
I^{\prime } &=&\int_{0}^{\infty }\frac{(n+1)\xi ^{\frac{\nu }{s}}}{%
z^{-(n+1)}e^{(n+1)\xi }-1}d\xi  \nonumber \\
&=&\int_{0}^{\infty }\frac{(n+1)\xi ^{\frac{\nu }{s}}}{e^{(n+1)(\xi -t)}-1}%
d\xi  \nonumber \\
&\simeq &-(n+1)\frac{t^{\frac{\nu }{s}+1}}{\frac{\nu }{s}+1}%
+2\sum_{j=1,3,5\cdots }\left( \matrix{ \frac{\nu }{s} \cr j}\right) t^{\frac{%
\nu }{s}-j}\frac{1}{(n+1)^{j}}\Gamma (1+j)\zeta (1+j).
\end{eqnarray}%
Then we obtain:
\begin{eqnarray}
h_{\sigma }(z) &=&\frac{1}{\Gamma (\sigma +1)}t^{\sigma }n\left[ 1+\frac{\pi
^{2}}{3}\frac{1}{n+1}\sigma (\sigma -1)t^{-2}\right.  \nonumber \\
&&\left. +\frac{\pi ^{4}}{45}\frac{1}{n}(1-\frac{1}{(n+1)^{3}})\sigma
(\sigma -1)(\sigma -2)(\sigma -3)t^{-4}+\cdots \right] .  \label{e20.2}
\end{eqnarray}%
Substituting Eq.(\ref{e20.2}) into Eq.(\ref{e18.2}), we obtain

\begin{eqnarray}
\frac{N}{V} &=&\frac{1}{\lambda ^{\nu }}\frac{2g}{\nu \Gamma (\frac{\nu }{2})%
}t^{\frac{\nu }{s}}n\left[ 1+\frac{\pi ^{2}}{3}\frac{1}{n+1}\frac{\nu }{s}(%
\frac{\nu }{s}-1)t^{-2}\right.  \nonumber \\
&&\left. +\frac{\pi ^{4}}{45}\frac{1}{n}(1-\frac{1}{(n+1)^{3}})\frac{\nu }{s}%
(\frac{\nu }{s}-1)(\frac{\nu }{s}-2)(\frac{\nu }{s}-3)t^{-4}+\cdots \right] .
\label{e20.3}
\end{eqnarray}

To obtain the thermodynamic functions for low temperatures and high
densities we first write down the expansion for the chemical potential from
Eq.(\ref{e20.3}):

\begin{equation}
\mu =\epsilon _{F}\left[ 1-\frac{\pi ^{2}}{3}\frac{1}{n+1}(\frac{\nu }{s}-1)(%
\frac{kT}{\epsilon _{F}})^{2}+\cdots \right] .  \label{e20.5}
\end{equation}%
The expansion parameter is $\frac{kT}{\epsilon _{F}}$. This expression
implies that when $\nu =s$ the chemical potential $\mu $ is equal to the
Fermi energy $\epsilon _{F}$ at an arbitrary temperature. A special case is
free particles in two dimensions, in which $\nu =s=2$. The relation between $%
\epsilon _{F}$ and the Fermi energy $\epsilon _{F}^{fermion}$ in Fermi-Dirac
statistics is

\begin{equation}
\epsilon _{F}=\epsilon _{F}^{fermion}\left( \frac{N-n}{Nn}\right) ^{\frac{s}{%
\nu }},
\end{equation}
where

\begin{equation}
\epsilon _{F}^{fermion}=\frac{\hbar ^{s}}{2m}\left[ \frac{2^{\nu }\pi ^{\nu
/2}}{g}\Gamma (\frac{\nu }{2}+1)\frac{N}{V}\right] ^{\frac{s}{\nu }}.
\end{equation}%
In other words, $\epsilon _{F}$ can be regarded as an analogue of the Fermi
energy in Gentile statistics, and $\epsilon _{F}$ will return to $\epsilon
_{F}^{fermion}$ when $n=1$.

With the help of Eq.(\ref{e20.5}), the asymptotic expansion of the internal
energy is then given by

\begin{equation}
\frac{U}{N}=\frac{N-n}{N}\frac{1}{\frac{s}{\nu }+1}\epsilon _{F}\left[ 1+%
\frac{\pi ^{2}}{3}\frac{1}{n+1}(\frac{\nu }{s}+1)(\frac{kT}{\epsilon _{F}}%
)^{2}+\cdots \right] .  \label{e20.8}
\end{equation}
The pressure of the gas is:

\begin{eqnarray}
P &=&\frac{1}{\frac{\nu }{s}+1}\frac{N-n}{V}\epsilon _{F}\left[ 1+\frac{\pi
^{2}}{3}\frac{1}{n+1}(\frac{\nu }{s}+1)(\frac{kT}{\epsilon _{F}})^{2}+\cdots %
\right]  \label{e20.9} \\
&&+\frac{n}{V}\epsilon _{F}\left[ 1-\frac{\pi ^{2}}{3}\frac{1}{n+1}(\frac{%
\nu }{s}-1)(\frac{kT}{\epsilon _{F}})^{2}+\cdots \right] .  \nonumber
\end{eqnarray}%
The free energy of the system follows directly from Eq.(\ref{e20.8}) and Eq.(%
\ref{e20.9})

\begin{equation}
\frac{F}{N}=\mu -\frac{PV}{N}=\frac{N-n}{N}\frac{1}{\frac{s}{\nu }+1}%
\epsilon _{F}\left[ 1-\frac{\pi ^{2}}{3}\frac{1}{n+1}(\frac{\nu }{s}+1)(%
\frac{kT}{\epsilon _{F}})^{2}+\cdots \right] ,
\end{equation}
whence we obtain the entropy of the system:

\begin{equation}
\frac{S}{Nk}=\frac{N-n}{N}\frac{2\pi ^{2}}{3}\frac{1}{n+1}\frac{\nu }{s}%
\frac{kT}{\epsilon _{F}}+\cdots .
\end{equation}%
From Eq.(\ref{e20.8}), the specific heat of the gas can be obtained

\begin{equation}
\frac{C_{V}}{Nk}=\frac{N-n}{N}\frac{2\pi ^{2}}{3}\frac{1}{n+1}\frac{\nu }{s}%
\frac{kT}{\epsilon _{F}}+\cdots .
\end{equation}
It is easy to see that there is a factor $(N-n)$ appearing in each
expression of the thermodynamic quantities, where $N$ is the total number of
particles and $n$ is the maximum occupation number of a state in Gentile
statistics. When $n$ is of the order of $N$, the values of the thermodynamic
quantities will be strongly suppressed by this factor. It is worth comparing
$C_{V}$ to the specific heat of the ideal Fermi gas $C_{V}^{fermion}$ so as
to discuss the difference between Gentile statistics and Fermi-Dirac
statistics:

\begin{equation}
C_{V}=\eta_{1} C_{V}^{fermion},
\end{equation}
where

\begin{equation}
\eta_{1} =\left( \frac{N-n}{N}\right) ^{1-\frac{s}{\nu }}\frac{2n^{\frac{s}{%
\nu }}}{n+1}
\end{equation}
Similarly, the magnetic susceptibility per unit volume of a system is given
by

\begin{equation}
\chi =\eta_{2} \chi ^{fermion},
\end{equation}%
where
\begin{equation}
\eta_{2}=\left( \frac{N-n}{Nn}\right) ^{1-\frac{s}{\nu }}
\end{equation}
and $\chi ^{fermion}$ is the magnetic susceptibility of a fermion system:

\begin{equation}
\chi ^{fermion}=\frac{g}{h^{\nu }}\frac{2\pi ^{\nu /2}}{s\Gamma (\frac{\nu }{%
2})}(2m)^{\frac{\nu }{s}}(\epsilon _{F}^{fermion})^{\frac{\nu }{s}-1}.
\end{equation}%
Obviously, the influence of the ground state depends sensitively on the
maximum occupation number $n$ when $\nu >s$.

\section*{V. Conclusions and outlook}

In summary, we first discuss the case $n\rightarrow N$. In some literature,
e.g., Ref. \cite{ter Haar}, the authors argue that such a case is just the
Bose case. In this paper we point out that this argument is valid only when
the fugacity $z<1$. When $z>1$, Gentile statistics with $n\rightarrow N$
does not return to Bose-Einstein statistics and in such a case the
contribution from ground state plays a very important role. Moreover, we
discuss the contribution of the ground state in Gentile statistics which is
ignored in the previous literature. The result shows that the ground state
contribution becomes important when the maximum occupation number $n$ is
large, especially in the case of low temperatures and high densities. The
results given by current literature focus only on the 3-dimensional space
with the dispersion relation $E=\frac{p^{2}}{2m}$. In the present paper, we
give the thermodynamic quantities for an arbitrary dispersion relation $E=%
\frac{p^{s}}{2m}$ in arbitrary $\nu $-dimensional space and find that the
result depends sensitively on the maximum occupation number $n$ when $\nu >s$%
.

As a generalization of Bose-Einstein and Fermi-Dirac statistics, fractional
statistics has been discussed for many years \cite{Leinaas Goldin,Wilczek}.
The particles obeying fractional statistics must not be real particles. As
an effective method, however, the introducing of such imaginary particles
can be used to deal with some complex interaction systems \cite{Canright}.
Generally speaking, there are two ways to achieve the fractional statistics:
(1) One is based on counting the number of many-body quantum states, i.e.,
generalizing the Pauli exclusion principle \cite{Gentile,Haldane,Wu}. The
most direct generalization is to allow more than one particles occupying one
state, and such an approach gives Gentile statistics \cite{Gentile}. Another
way is to introduce a parameter $\alpha $ valued from $0$ to $1$ in the
expression of the number of quantum states \cite{Wu}. Bose-Einstein or
Fermi-Dirac statistics becomes its limit case when $\alpha $ is $0$ or $1$,
respectively. (2) Alternatively, the fractional statistics can be achieved
by analyzing the symmetry properties of the wave function. As is well known,
the wave function will change a phase factor when two identical particles
exchanges. The phase factor can be $+1$ or $-1$ related to bosons or
fermions, respectively. When this result is generalized to an arbitrary
phase factor $e^{i\theta }$, the concept of anyon is obtained \cite{Wilczek}%
. Comparing these two approaches to fractional statistics a question arises
naturally: Are there any connections between them? To answer this question
we need to build a bridge between the generalized Pauli principle (in which
the maximum occupation number is extended to an arbitrary integer $n$) and
the exchange symmetry of identical particles. We will discuss this problem
elsewhere \cite{Ours}.

\vskip0.5cm

We would like to thank Dr. Yong Liu for sending us some important
references, and we are very indebted to Dr. G. Zeitrauman for his
encouragement. This work is supported in part by LiuHui fund and the Science
fund of Tianjin University, P. R. China.

\end{document}